# Neutrino Rocket Jet Model: An Explanation of High-velocity Pulsars and Their Spin-down Evolution

Zheng Li[1,2,3,4,5,6], Qiu-He Peng[5], Miao Kang[7], Xiang Liu[1,2,3,4,6], Ming Zhang[1,2,3,4,6], Yong-Feng Huang[5,8], and Chih-Kang Chou[9]

[1] Xinjiang Astronomical Observatory, Chinese Academy of Sciences, 150 Science 1-Street, Urumqi 830011, People's Republic of China
[2] Key Laboratory for Radio Astronomy, Chinese Academy of Sciences, 2 West Beijing Road, Nanjing 210008, People's Republic of China
[3] University of Chinese Academy of Sciences, 19A Yuquan Road, Beijing 100049, People's Republic of China
[4] College of Big Data and Information Engineering, Guizhou University, Guiyang 550025, People's Republic of China
[5] School of Astronomy and Space Science, Nanjing University, Nanjing 210023, People's Republic of China; qhpeng@nju.edu.cn
[6] Key Laboratory of Radio Astrophysics in Xinjiang Province, Urumqi 830011, People's Republic of China
[7] School of Physics and Electronics, Henan University, Kaifeng, People's Republic of China
[8] Key Laboratory of Modern Astronomy and Astrophysics (Nanjing University), Ministry of Education, Nanjing 210023, People's Republic of China
[9] National Astronomical Observatories, Chinese Academy of Sciences, Beijing, People's Republic of China
Received 2022 February 21; revised 2022 May 3; accepted 2022 May 3; published 2022 June 2

## Abstract

The fact that the spatial velocity of pulsars is generally higher than that of their progenitor stars has bothered astronomers for nearly 50 years. It has been extensively argued that the high pulsar velocity should be acquired during a natal kick process on a timescale of 100 ms–10 s in the supernova explosion, in which some asymmetrical dynamical mechanism plays a key role. However, a satisfactory picture generally is still lacking. In this study, it is argued that the neutrino rocket model can well account for the high speed as well as the long-term evolution behaviors of pulsars. The neutrinos are emitted from superfluid vortex neutrons through the neutrino cyclotron radiation mechanism. The unique characters of left-handed neutrinos and right-handed antineutrinos resulting from the nonconservation of parity in weak interactions play a major role in the spatial asymmetry. The continuous acceleration of pulsars can be naturally explained by this model, which yields a maximum velocity surpassing 1000 km s$^{-1}$. The alignment between the spinning axis and the direction of motion observed for the Crab pulsar (PSR 0531) and the Vela pulsar (PSR 0833) can be well accounted for. The observed correlation between the spin-down rate and the period of long-period pulsars with $P \gtrsim 0.5$ s can also be satisfactorily explained.

*Unified Astronomy Thesaurus concepts:* Neutron stars (1108); Pulsars (1306); Proper motions (1295); Stellar evolutionary models (2046)

## 1. Introduction

Ever since their discovery in 1967, pulsars still remain mysterious in many aspects. For example, the internal composition of pulsars is highly disputed (Geng et al. 2021). Another interesting issue concerns their spatial motions. Astronomers are amazed to find that the spatial velocity of pulsars is generally high (Gunn & Ostriker 1970; Minkowski 1970; Lyne & Lorimer 1994). The average velocity of pulsars is about 450 km s$^{-1}$, which is much larger than that of the progenitors of supernova explosions and other Population I stars ($\sim$30 km s$^{-1}$). Five pulsars are even observed to have a velocity exceeding 1000 km s$^{-1}$. For example, PSR B2224+65 in the Guitar nebula is found to be associated with a bow shock (Cordes et al. 1993), which hints at a velocity of >1000 km s$^{-1}$.

Interestingly, for at least two famous young pulsars, namely, the crab pulsar (PSR 0531) and the Vela pulsar (PSR 0833), the direction of motion is found to be roughly parallel to the pulsar spin axis (Lai et al. 2001). It indicates that these neutron stars may have experienced an effective kick along the direction of the spinning axis at birth. A continuous dynamical acceleration mechanism acting along the direction of the spinning axis is needed. However, the exact acceleration mechanism has troubled astronomers for nearly 50 years.

Until now, most of the theories proposed to solve this problem are based on some specially asymmetrical physical mechanism during the process of supernovae explosions (on a timescale of 100 ms–10 s). These short-duration dynamic kicks are called the birth kick mechanisms. For example, one kind of model is based on the asymmetrical matter ejection or asymmetrical neutrino emission caused by dynamical instabilities of fluids due to stellar rotation and magnetic fields in the supernovae cores during the gravitational collapse (Burrows & Fryxell 1992; Herant et al. 1994; Janka & Mueller 1994; Burrows et al. 1995). A second kind of natal kick mechanisms resorts to the superstrong magnetic field of $\sim 10^{16}$ G, which can cause significant asymmetrical scattering of neutrinos on other particles. Due to the nonconservation of parity in weak interactions, asymmetrical neutrino emission is possible during these scatters, which leads to a high rebound speed for the nascent neutron star (Horowitz & Li 1998; Lai & Qian 1998; Arras & Lai 1999). Third, neutrino oscillations in a superstrong magnetic field ($10^{15}$–$10^{16}$ G) may also result in a high kick speed (Kusenko & Segrè 1996; Grasso 1998). Finally, there is another kick mechanism named the electromagnetic rocket model (Harrison & Tademaru 1975; Lai et al. 2001; Lai 2004; Xu et al. 2022). In this model, a rotating off-centered magnetic dipole can generate asymmetrical electromagnetic radiation, which in turn may generate a recoil velocity (up to 1000 km s$^{-1}$) along the spinning axis. The kick energy comes from the pulsar's rotational energy, which spins down on a timescale of $\tau_{\rm em} \simeq B_{13}^{-2}(\nu/1\,{\rm kHz})^{-2}$ yr, where $B_{13}$ is the pulsar magnetic field in units of $10^{13}$ G and $\nu$ is its spin frequency.

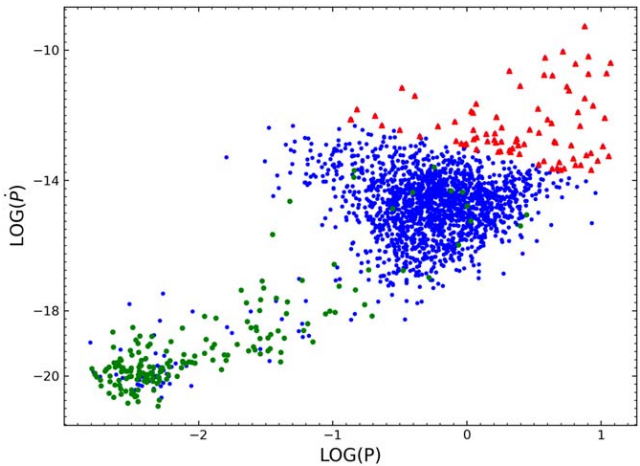

**Figure 1.** The spin-down rate of the observed pulsars, $\dot{P}$, vs. their spin period, $P$. The dots in blue, green, and red stand for normal pulsars, pulsars in binary systems, and magnetars with a magnetic field $B > 10^{13}$ G, respectively. The observational data are taken from Hobbs et al. (2004), with a catalog version of 1.65.

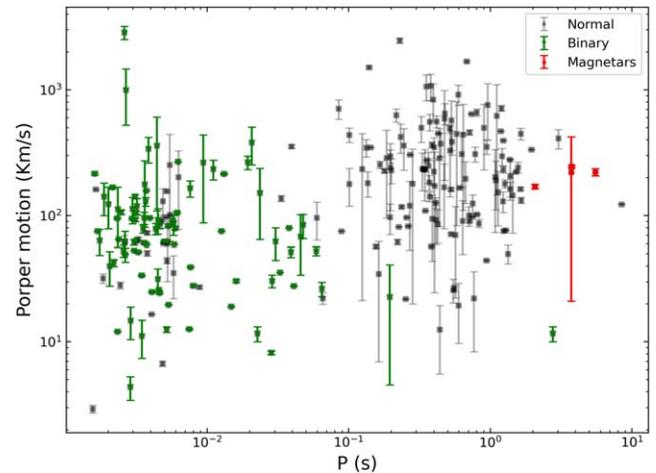

**Figure 2.** Proper motion vs. period for observed pulsars. Gray points represent normal pulsars, red points represent magnetars ($B > 10^{13}$ G), and the green dots stand for pulsars in binary systems The observational data are taken from Hobbs et al. (2004) and the catalog version is 1.65.

However, all these theories have some difficulties. For example, for the first kind of model, how the asymmetrical matter ejection or neutrino emission can be acquired is still largely uncertain. For the other theories, a magnetic field up to $10^{15}$–$10^{16}$ G is generally required. But actually, most pulsars with a velocity exceeding 300 km s$^{-1}$ have a magnetic field less than $5 \times 10^{12}$ G.

It is interesting to note that neutrino cyclotron emission that is produced by neutron superfluid vortices inside neutron stars can act as another mechanism for neutron stars spinning down (Peng et al. 1982). It may also lead to a continuous acceleration of the pulsar, resulting in a kick velocity up to 1000 km s$^{-1}$ (Peng et al. 2004). Note that this is not a birth kick mechanism as described in the previous paragraph. On the contrary, it acts over a long time and is effective for long-period pulsars. In this kick mechanism, the unique characteristics of left-handed neutrinos and right-handed antineutrinos due to the nonconservation of parity in weak interactions are considered. It can naturally explain the alignment of the spinning axes and the kick velocity observed in the Crab pulsar (PSR 0531) and the Vela pulsar (PSR 0833). In this study, we present detailed calculations in the framework of this mechanism and compare the results with observations.

Our paper is organized as follows. In Section 2, the specific process of data processing is introduced. The basic idea of the neutrino rocket jet model is explained in Section 3. Finally, Section 4 presents our discussion.

## 2. Sample and Data Processing

The pulsar data we used is taken from Hobbs et al. (2004), and the catalog version is 1.65. There are 2492 pulsars with necessary data available in our sample. Figure 1 shows these pulsars on the $P$–$\dot{P}$ diagram. To investigate whether the pulsars have a gradual acceleration trend, we have plotted their available observed velocities in the catalog, which are only for 329 pulsars, versus their period in Figure 2. After that, the pulsars have been divided into five groups according to their velocities, i.e., 0–100 km s$^{-1}$ group, 100–300 km s$^{-1}$ group, 300–500 km s$^{-1}$ group, 500–1000 km s$^{-1}$ group, and $\geqslant$1000 km s$^{-1}$ group. Each velocity measurement has an error bar. To decide which group each pulsar belongs to, we need to consider the effect of the error bar. For a particular speed range, there could be four kinds of error bars. Figure 3 shows the four situations of the error bars for the speed range of 0–100 km s$^{-1}$. The cases are similar for the other four speed ranges.

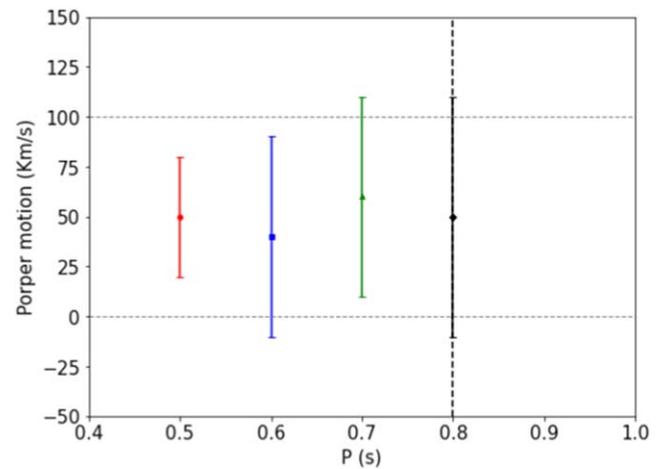

**Figure 3.** A schematic illustration of the four kinds of error bars for pulsars with a speed range of 0–100 km s$^{-1}$. The four kinds of error bars are also applicable to the other speed ranges.

To get the probability for each error bar to fall in a particular speed range, the probability density function of the Gaussian distribution $f(x) = \frac{1}{\sqrt{2\pi}\sigma} \exp\left(-\frac{(x-\mu)^2}{2\sigma^2}\right)$ has been considered. For example, for the speed range of 0–100 km s$^{-1}$, we first examine whether a pulsar's proper motion speed falls within this speed range. If it does, we then calculate the probability of its error bar, which is just the probability density that the error bar falls within this range for more than $1\sigma$. It is found that the $3\sigma$ confidence level requires a probability larger than 99.73%. Therefore, we use this percentage value as a criterion to judge whether the pulsar belongs to the 0–100 km s$^{-1}$ group or not. Cases are similar for the other four speed ranges in judging the actual sample probability. Figure 4 is a schematic illustration of the probability density for the four kinds of error bars in the 0–100 km s$^{-1}$ speed range. Using the above method, the number of pulsar members of each speed group have been determined. The final results are summarized in Table 1. The percentage of pulsars with a proper motion larger than 300 km s$^{-1}$ for each of the five period ranges has also been calculated and presented in Table 1.

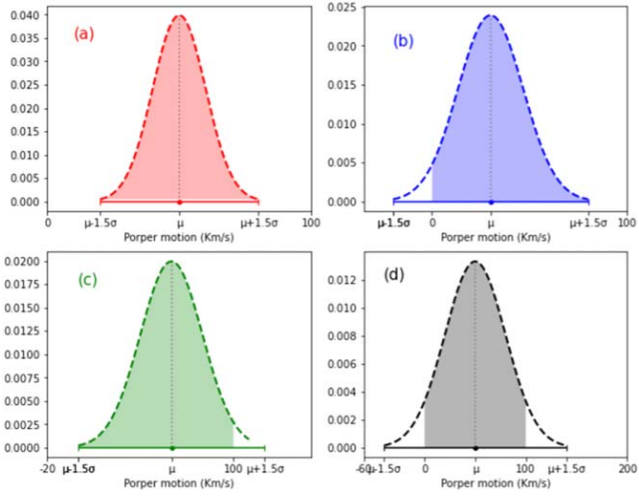

**Figure 4.** The probability density diagram for the four kinds of error bars in the 0–100 km s$^{-1}$ speed range. Cases are similar for the other speed ranges.

From the statistics in Table 1, a clear tendency could be seen that the percentage of high-speed pulsars (>300 km s$^{-1}$) increases with the increasing period, showing a long-term acceleration behavior. Nevertheless, some millisecond pulsars are found to have a high speed. These high-speed millisecond pulsars, with $P < 0.1$ s and $V > 300$ km s$^{-1}$, are usually in binary systems. They are believed to have experienced a magnetic field decay stage caused by the Ohm dissipation process. Their spinning has also been accelerated due to long-term accretion. So, they will not follow the period–velocity relation predicted by our following model.

To illustrate the long-term evolution of pulsars more clearly, we now concentrate on the pulsars with a period $P > 0.5$ s and study their distribution on the $P$–$\dot{P}$ diagram. Since there is a clear positive correlation between $P$ and $\dot{P}$, we fit the observational data points by using a linear function. Generally speaking, the linear fitting method requires that the random error bars satisfy the Gaussian distribution. However, this requirement is not satisfied all the time. A maximum likelihood function method based on the Markov Chain Monte Carlo (MCMC) simulations is thus used to derive the likelihood in these cases. (Hogg et al. 2010). It is a better solution when the distribution of error bars does not satisfy the Gaussian function. Here the likelihood function is

$$\ln p(y|x, \sigma, m, b, f) = -\frac{1}{2}\sum_n \left[\frac{(y_n - mx_n - b)^2}{s_n^2} + \ln(2\pi s_n^2)\right], \quad (1)$$

where $s_n^2 = \sigma_n^2 + f^2(mx_n + b)^2$ is an estimate of the mean absolute uncertainty, $m$ and $b$ are the coefficients of the linear function, $f$ is the intrinsic Gaussian variance.

Today, more than 2600 pulsars have been observed, about half of which have a period larger than 0.5 s. Some pulsars even have a period $P > 10$ s. After running 5000 steps of MCMC on these long-period pulsars ($P > 0.5$ s), the maximum likelihood results for the parameters of the linear function are derived, which are plotted in Figure 5. Figure 6 further shows the best linear fit result. It could be seen that for pulsars with $P > 0.5$ s, the $P$–$\dot{P}$ diagram is best fitted by a straight line with a slope of $+2$, i.e., $\dot{P} \propto P^2$. This result is contrary to the prediction of the standard classic pulsar model. Soon after the discovery of pulsars in 1967, Ostriker and Gunn proposed the famous "rotating magnetic dipole radiation mechanism" to explain their spin down (Ostriker & Gunn 1969). According to this popular model, the spin-down rate is inversely proportional to the spin period, i.e., $\dot{P} \propto P^{-1}$, so that the distribution of pulsars on the logarithmic $P$–$\dot{P}$ diagram should roughly follow a straight line with a slope of $-1$. However, Figure 6 clearly shows that the observed pulsars do not follow the $\dot{P} \propto P^{-1}$ relation. On the contrary, it is $\dot{P} \propto P^2$. It is interesting to note that such a behavior is exactly consistent with the relation predicted by the neutrino rocket jet model (Peng et al. 1982). According to this theory, the spin down of pulsars is due to the neutrino radiation of the neutron superfluid vortex.

**Table 1**
Number of Pulsars Falling in Particular Velocity($V$)–Period($P$) Ranges Based on $3\sigma$ Statistics

| | $P$ (s) | | | | |
|---|---|---|---|---|---|
| $V$ (km s$^{-1}$) | <0.1 | 0.1 ∼ 0.3 | 0.3 ∼ 0.5 | 0.5 ∼ 1 | >1 |
| <100 | 64 | 6 | 1 | 5 | 2 |
| 100 ∼ 300 | 18 | 10 | 9 | 8 | 10 |
| 300 ∼ 500 | 5 | 3 | 3 | 0 | 5 |
| 500 ∼ 1000 | 1 | 0 | 0 | 5 | 3 |
| >1000 | 1 | 1 | 2 | 1 | 1 |
| Percentage of >300 | 7.87% | 20% | 33.33% | 31.58% | 42.86% |

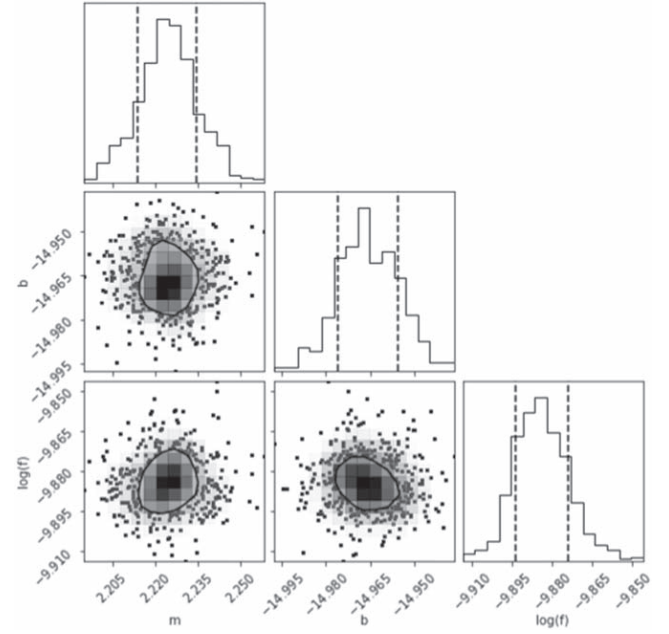

**Figure 5.** A corner plot showing the probability distribution of the three parameters of the linear function engaged to fit the $P$–$\dot{P}$ data points. The dashed lines illustrate the $1\sigma$ range of the parameters.

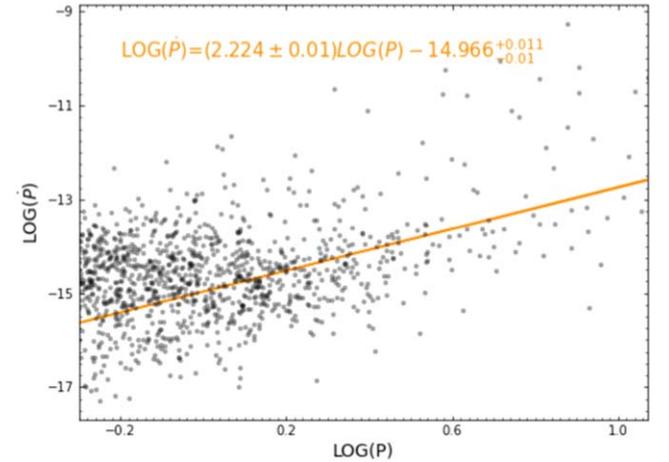

**Figure 6.** The best linear fitting result for long-period pulsars with $P \gtrsim 0.5$ s.

### 3. Basic Idea of the Neutrino Rocket Jet Model

#### 3.1. Neutrino Emission and the Spinning Down of Pulsars

According to the Weinberg–Salam electroweak unification theory, neutrons in a circular motion can emit neutrino–antineutrino pairs via the neutral current in the electroweak interactions. Similarly, the superfluid vortex neutrons inside neutron stars, which includes the $^1S_0$ isotropic superfluid region and the $^3P_2$ anisotropic superfluid region, can also continuously emit neutrinos and antineutrinos. This process is referred to as neutrino cyclotron radiation. The neutrino pairs will directly escape the neutron star, leading to a continuous energy loss and spin down of the compact star. Neutron stars formed during the violent supernova explosion are in a highly turbulent state. The resulting neutron superfluid vortex must be in a very high quantum state. Therefore, the neutrinos emitted by these neutron superfluid vortices cannot be ignored.

The interaction between electrons, magnetic moments, and the odd neutron magnetic moment determines the coupling of the neutron superfluid with the neutron star crust and the entire magnetosphere (Sauls 1989). Moreover, the Ekman pump mechanism connected to the superfluid vortex motion also contributes to the coupling. Through this connection, the energy consumed due to the emission of neutrinos from $^1S_0$ and $^3P_2$ superfluid is supplied by the rotational kinetic energy of the whole neutron star, which increases its spin period. Therefore the emission of neutrinos can lead to the spin down of pulsars, which is an entirely new mechanism significantly different from the standard magnetic dipole radiation model (Peng et al. 1982). It should be noted that for short-period pulsars, the spin down is still mainly due to the standard magnetic dipole radiation mechanism. But for long-period pulsars, the spin down will be dominated by the emission of neutrinos from neutron superfluid vortices. The $P$–$\dot{P}$ diagram plotted in Figure 6 strongly supports this mechanism.

#### 3.2. Neutron Superfluid Vortex Motion in Neutron Stars

Neutron star recoils due to neutrino radiation from the neutron superfluid vortices are studied here. For the sake of completeness, the internal structure of neutron stars will be briefly introduced first. A neutron star has a thin crust that includes the isotropic $^1S_0$ neutron superfluid region and the anisotropic $^3P_2$ neutron superfluid region. In the isotropic $^1S_0$ neutron superfluid region, the density is $10^{11}$ g cm$^{-3} < \rho < 1.4 \times 10^{14}$ g cm$^{-3}$. In the anisotropic $^3P_2$ neutron superfluid region, the density is $3.3 \times 10^{14}$ g cm$^{-3} < \rho < 5.2 \times 10^{14}$ g cm$^{-3}$ (note that the nuclear saturation density is $2.8 \times 10^{14}$ g cm$^{-3}$).

Rotating superfluids are quantized to become superfluid vortex flows or eddy currents or whirling fluids (Feynman & Cohen 1955), analogous to type I superconductivity. The neutron superfluid inside neutron stars is in a vortex state. That is, there are a large number of vortex lines, called the vortex filaments. Usually, these vortex filaments are arranged in a periodic lattice parallel to the neutron star's rotation axis. They rotate almost rigidly around the rotation axis. The circulation of each vortex filament (the intensity, $\Gamma$) is quantized as

$$\Gamma = \oint \boldsymbol{v} \cdot d\boldsymbol{l} = n\Gamma_0, \; \Gamma_0 = \frac{2\pi\hbar}{2m_n}, \qquad (2)$$

where $n$ is the circulation quantum number of the vortex, $m_n$ is the mass of neutron, $\hbar$ is Planck's constant divided by $2\pi$, and $\Gamma_0$ is the unit vortex quantum (Feynman & Cohen 1955).

The supercurrent vortex's core is assumed to be cylindrical and the normal neutron fluid is immersed in the supercurrent neutron sea. The physical reasons for this assumption will be explained below. The core radius of the vortex ($a_0$) can be estimated as follows. In the core of the vortex, the position uncertainty of neutrons is $\Delta x \sim a_0$. The momentum uncertainty of neutrons is $\Delta p \sim \hbar/a_0$ according to the Heisenberg's principle of uncertainty, and then the uncertainty of neutron energy is $\Delta E \sim \Delta P^2/2m_n \sim \hbar^2/2m_n a_0^2$. The neutron fluid in the core of the vortex is in the normal state (no Cooper pairs are formed), where the energy uncertainty of neutrons is larger than the energy gap of neutron superfluid and the binding energy of Cooper pairs, i.e., $\Delta E > \Delta_n$. Thus we have

$$a_0 \sim \hbar/\sqrt{2m_n \Delta_n}. \qquad (3)$$

For the vortices of isotropic neutron superfluids, $\Delta_n(^1S_0) \approx 2$ MeV, $a_0 \sim 10^{-12}$ cm, and for the vortices of anisotropic neutron superfluids, $\Delta_n(^3P_2) \approx 0.045$ MeV, $a_0 \sim 10^{-11}$ cm.

Outside the core of the vortex, neutrons are in a superfluid state. The superfluid neutrons revolve around the vortex line with a velocity of

$$v_s(r) = \frac{n\hbar}{2m_n r}, \qquad (4)$$

where $r$ is the distance from the axis of the vortex filament. The relativistic effect on the masses of neutrons should be considered because the rotating velocity around the vortex axes may be near the speed of light when the circulation quantum number of the vortex is very high ($n \approx 10^2$–$10^4$). So, we have $m_n = \frac{m_n^{(0)}}{\sqrt{1-(v/c)^2}}$, where $m_n^{(0)}$ is the rest mass of neutrons.

The angular velocity of the neutron superfluid revolving around the vortex filament is

$$\omega_s(r) = \frac{n\hbar}{2m_n r^2}. \qquad (5)$$

Therefore, the rotation of superfluid neutrons around the vortex filament is in a differential state. In the vicinity of $r \sim a_0$, the velocity and angular velocity of superfluid neutrons rotating around the vortex axis reach the maximum values, which are

$$v_{s,\max} = \frac{n\hbar}{2m_n a_0}, \quad (6)$$

$$\omega_{s,\max} = \frac{n\hbar}{2m_n a_0^2}. \quad (7)$$

We have $\omega_{s,\max}(^1S_0) \sim 10^{21} n$ s$^{-1}$ and $\omega_{s,\max}(^3P_2) \sim 10^{19} n$ s$^{-1}$. Inside the core of the vortexes ($r < a_0$), however, the normal neutron fluid revolves rigidly at an angular velocity of $\omega_{s,\max}$.

According to Feynman & Cohen (1955), the number of superfluid vortex filaments per unit area is $2\Omega/n\Gamma_0$. Then the average distance ($b$) between two neighboring vortex filaments and the total number of superfluid vertices ($N_{\text{vertice}}$) in the superfluid region are respectively,

$$b = \left(\frac{\bar{n}\hbar}{2m_n\Omega}\right)^{1/2}, \quad (8)$$

$$N_{\text{vertice}} = \frac{2m_n\Omega}{n\hbar}R_s^2, \quad (9)$$

where $\Omega$ is the mean angular velocity of rotation, $R_s$ is the radius of the superfluid region, and $\bar{n}$ is the circulation quantum number of each vortex filament on average.

In general, the cyclic quantum numbers of the liquid $^4$He and $^3$He vortex filaments are very small at low temperatures and may even be $n = 1$. However, the interior of a nascent neutron star must be in a turbulent vortex state, because it originated from the core collapse during a violent supernova explosion that lasted for less than 10 s. It is hard to transport the angular momentum of the collapsed core outwards. As a result, a considerable portion of the stellar angular momentum is stored in the highly turbulent neutron fluid (Huang et al. 1982). The classical circulation of vortex filaments, intensity $\Gamma$, may be extensive. As the internal temperature of the neutron star drops below the corresponding critical temperature, $^1S_0$ and $^3P_2$ superfluid states will appear in succession. Note that the $^1S_0$ superfluid state is isotropic, while the $^3P_2$ state is anisotropic.

When the $^3P_2$ state becomes anisotropic, the violent classical turbulent vortex state will have a large vortex quantum number. It is expected that the quantum number $n$ in Equation (4) may reach $10^2$–$10^4$ in the superfluid region for young pulsars.

In the superfluid vortex, a neutron rotating around the vortex center with angular velocity $\Omega$ can emit neutrino pairs with a power of (Peng et al. 1982)

$$p(\omega, r) = A\omega^8 r^2, \quad (10)$$

where the coefficient is $A = 4.7 \times 10^{-159}$ erg s$^7$ cm$^{-2}$. In this study, the total neutrino radiation power from the superfluid neutrons rotating in a single superfluid vortex and from the typical neutrons spinning in the vortex center is calculated. It is derived as

$$w_1 = \frac{\pi^4}{3}cA\frac{\omega_{s,\max}^7 a_0^6}{m_n^2}\rho_n^2 R, \quad (11)$$

where $\rho_n$ is the mass density of the neutron fluid at the core of the vortex.

Although the radiation coefficient $A$ is small in Equation (10), the radiation power is proportional to the eighth power of the angular velocity of the neutrons around the superfluid vortex axis. The angular velocity $\omega$ itself may be as large as $10^{22}$ s$^{-1}$, which can enhance the radiation power to a large value. As a result, the neutrino cyclotron radiation by superfluid vortex neutrons can effectively lead to a secular spin down of long-period pulsars (Peng et al. 1982). According to Equations (7), (9), and (11), the total neutrino radiation power due to the whole neutron superfluid vortex body is

$$W_v = \frac{\pi^5}{48}\frac{cAh^6}{(m_n a_0)^8}\frac{\bar{n^7}}{\bar{n}}\rho^2 R^3 P^{-1}. \quad (12)$$

To estimate the superfluid vortex core radius $a_0$, Equation (6) of Ruderman (1976) can be used, which is

$$a_0 = E_F/k_F\Delta \approx (3\pi^2)^{1/3}\frac{\hbar^2}{2m_n^{4/3}}\frac{\rho^{1/3}}{\Delta}$$

$$\approx 2.7\left(\frac{\Delta_S}{2.5 \text{ MeV}}\right)^{-1}\left(\frac{\rho}{2 \times 10^{12} \text{ g cm}^{-3}}\right)^{1/3} \text{ fm}. \quad (13)$$

Equation (12) can then be simplified as

$$W_v \approx B_S G(n) P^{-1}, \quad (14)$$

where

$$B_S = 1.6 \times 10^{29}\left(\frac{\Delta_S}{2.5 \text{ MeV}}\right)^8\left(\frac{\rho}{2 \times 10^{12} \text{ g cm}^{-3}}\right)^{-2/3} R_{10}^3, \quad (15)$$

$$G(n) = \frac{\bar{n^7}}{\bar{n}}. \quad (16)$$

The total neutrino radiation power of the anisotropic $^3P_2$ neutron superfluid vortex is

$$W_v^{(P)} \approx B_P G_P(n_P) P^{-1}, \quad (17)$$

with $B_P \approx 1.0 \times 10^{16}\left(\frac{\Delta_P}{0.045 \text{ MeV}}\right)^8\left(\frac{\rho}{\rho_{\text{nuc}}}\right)^2\left(\frac{R_P}{5 \text{ km}}\right)^3$. Here $\rho_{\text{nuc}} = 2.8 \times 10^{14}$ g cm$^{-3}$ is the nuclear saturation density, and $R_P$ is the radius of the $^3P_2$ superfluid region. This neutrino radiation power is much lower than the full neutrino radiation power from the isotropic $^1S_0$ neutron superfluid vortex and hence can be neglected. Therefore, only the neutrino radiation power from isotropic $^1S_0$ neutron superfluid vortex needs to be calculated.

The superfluid vortex quantum number is not a constant, but evolves with time. The quantum number $G(n)$ is initially very high when the neutron star is born. Then, with the increase of the pulsar age, the spin period gradually increases and $G(n)$ gradually decreases, which may be associated with pulsar glitches.

In reality, both the standard magnetic dipole radiation and the neutrino radiation from the neutron superfluid vortices may cause the pulsar to spin down. The overall spin-down rate can be calculated as (Peng et al. 1982)

$$\dot{P} = A_\gamma e^{-t/\tau_D} P^{-1} + \frac{B_S}{4\pi^2 I}G(n) P^2, \quad (18)$$

$$A_\gamma = \frac{2\pi^2}{3}\frac{B^2 R^6 \sin^2\alpha}{c^3 I}. \quad (19)$$

In Equation (18), the first term corresponds to the effect of standard magnetic dipole radiation, and the second term represents the effect of neutrino radiation from neutron superfluid vortices. Here $I$ denotes the moment of inertia of the neutron star and $R_{10}$ is the neutron star radius in units of 10 km. $B$ denotes the surface magnetic field strength and $\alpha$ is the angle between the magnetic field and the spin axis. The factor of $e^{t/\tau_D}$ is introduced to depict the decay of the magnetic field. But here it can be taken as $e^{t/\tau_D} \approx 1$ since the timescale of magnetic decay ($\tau_D \approx 10^7$ yr) is very long compared with the timescale of neutron star acceleration ($\ll 10^6$ yr). Then from Equation (18), we obtain

$$\dot{P} = A_\gamma P^{-1} + C(n) P^2, \quad (20)$$

$$C(n) \approx \frac{B_S}{4\pi^2 I} G(n). \quad (21)$$

The first term in Equation (20) is dominative only for a short time after the birth of the pulsar.

### 3.3. Neutrino Rocket Jet Model Engaging Neutron Superfluid Vortices

#### 3.3.1. Basic Idea

The nonconservation of parity in weak interactions leads to the left–right asymmetry of neutrinos. Neutrons in a circular motion and the flux of neutrinos emitted by a neutron superfluid vortex possess axial symmetry with respect to the spin axis. However, perpendicular to the spin axis (i.e., the equator), they have up–down asymmetry. In the $^3P_2$ neutron superfluid region, the anomalous magnetic moment of the Cooper pairs in a strong magnetic field will be directed opposite to the magnetic field, because the magnetic gyro ratio of neutrons is negative. This directional asymmetry can lead to an asymmetry of the neutrinos emitted by the neutron superfluid vortex.

The momentum of the emitted left-handed neutrinos is opposite to the pulsar spin vector, and these neutrinos will also take away some of the angular momentum correspondingly. So, the angular momentum of the neutron star will gradually decrease, naturally leading to a spin down. The spin of a neutrino is opposite to its momentum (correspondingly, the antineutrino spin is parallel to its momentum), which means the momentum of the neutrino flux must be opposite to the spin vector of the neutron star. Consequently, the neutron star will obtain a recoil velocity opposite to the moment of the emitted neutrinos. The momentum of the antineutrinos emitted is also opposite to the spin of the neutron star by a similar argument. So, the recoil caused by both neutrinos and antineutrinos accelerates the pulsar in the direction parallel to its spin vector. In other words, the pulsar will be gradually accelerated in the direction of its spin due to the continuous emission of neutrinos by superfluid vortices, which should also be associated with a secular spin down. This is effectively a neutrino rocket jet mechanism, through which the neutron star can obtain a very high recoil velocity along its spin axis.

#### 3.3.2. Recoil Speed

The recoil force exerted on the neutron star due to the emitted neutrinos and antineutrinos is

$$\frac{dp_v}{dt} = \frac{W_v}{c}. \quad (22)$$

The growth rate of the neutron star momentum is then

$$M \frac{dV}{dt} = \frac{W_v}{c} = \frac{B_S}{cP} G(n), \quad (23)$$

or equivalently

$$\frac{dV}{dP} = \frac{B_S}{Mc} \frac{G(n)}{P\dot{P}}, \quad (24)$$

where $M$ is the mass of the neutron star. Substituting Equation (20) into Equation (24), we obtain

$$\frac{dV}{dP} = DG(n), \quad (25)$$

$$D = \frac{B_S}{A_\gamma Mc} \approx 111 \left(\frac{B}{10^{12}\,\mathrm{G}}\right)^{-2} R_{10}^{-3} I_{45} \left(\frac{\sin^2\alpha}{0.1}\right)^{-1} \quad (26)$$
$$\times \left(\frac{(\Delta_S/2.5\,\mathrm{MeV})^8}{(\rho_S/2\times 10^{12}\,\mathrm{g\,cm^{-3}})^{2/3}}\right).$$

Finally, the neutron star recoil velocity due to the rocket jet effect is

$$V(P) = V_0 + D \int_{P_0}^{P} G(n) dP. \quad (27)$$

As mentioned before, the superfluid vortex quantum number is very high for the neutron stars at birth. It may decrease as the pulsar period gradually increases. A simple expression is adopted here,

$$G(n) \equiv G(P) = G^{(0)} \left(\frac{P}{P_0}\right)^{-\beta}, \quad (28)$$

where $P_0$ is the initial period of the pulsar and $G^{(0)}$ is the initial value of $G$. Usually, the decay power index $\beta$ is less than 1. Equation (27) then gives

$$V(P) = V_0 + \frac{1}{1-\beta} D P_0 G^{(0)} (P/P_0)^{(1-\beta)}. \quad (29)$$

In this study, some typical parameters are taken as $P_0 = 1$ ms, $G^{(0)} = 1.0 \times 10^6$, and $V_0 = 30$ km s$^{-1}$. For the decay index ($\beta$), several different values are assumed. The evolution of pulsar speed can then be calculated and compared with observations. The results are plotted in Figure 7(a). It is found that when the period is 0.3 s, the velocity of the star can increase to 76 km s$^{-1}$, 114 km s$^{-1}$, and 162 km s$^{-1}$ for $\beta = 0.5$, 0.3, and 0.2, respectively. When $\beta$ increases, the recoil velocity decreases. The results for $G^{(0)} = 1.0 \times 10^7$ are also presented in Figure 7(b). It is found that the velocity (at the moment of $P = 0.3$ s) will correspondingly be 392 km s$^{-1}$, 874 km s$^{-1}$, and 1346 km s$^{-1}$ for $\beta = 0.5$, 0.3, and 0.2, respectively. Obviously, a larger $G^{(0)}$ will make the acceleration of the pulsar more efficient. Note that many millisecond pulsars have a high speed exceeding the expectation of this model. As mentioned before, they are actually in binary systems and may have been affected by binary evolution.

In the above calculations, the magnetic field is taken as $B = 10^{12}$ G for simplicity. However, note that different pulsars may have different magnetic field strengths, as shown in Figure 8. Especially, for those millisecond pulsars with $P < 0.01$ s in Figure 7(b), the magnetic field is typically in a range of $10^8$–$10^9$ G and is much lower than this value.

### 4. Discussion

The idea of a neutron superfluid vortex is very similar to that of the liquid $^4$He superfluid vortex proposed by R. P. Feynman (Feynman & Cohen 1955). It is interesting to see that such a superfluid vortex state could play an important role inside neutron stars. The idea that neutrons in rotational motion can radiate neutrinos, and the idea that the neutron superfluid vortex motion inside neutron stars can be quantized, are effectively combined in this study. The origin of high-speed pulsars and their distribution on the $P$–$\dot{P}$ diagram can be well explained in this framework. In general, in condensed matter physics, the Bogoliubov–De Gennes equation describes a system with a vortex at the vanishing temperature. But this equation involves multibody interactions and it is difficult to apply it to the interior of neutron stars. In this study, R. P. Feynman's idea is extended to explore the superfluid vortex inside neutron stars from a new aspect.

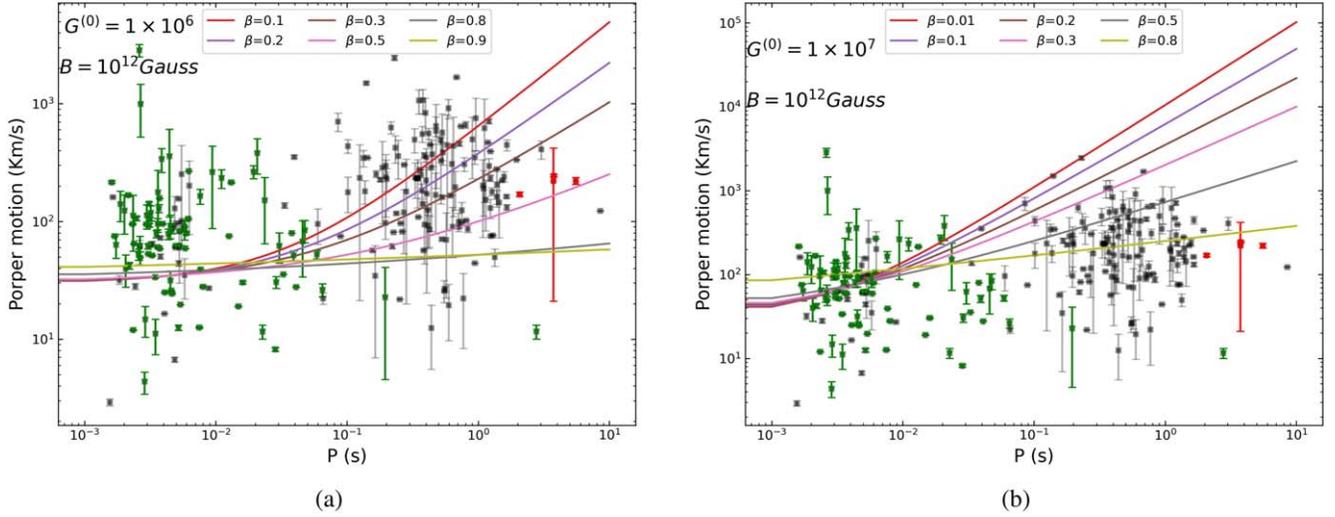

**Figure 7.** Pulsar velocity vs. its spin period. The points correspond to observational data, and the lines are theoretical results from the neutrino rocket jet model with (a) $G^{(0)} = 1 \times 10^6$ and (b) $G^{(0)} = 1 \times 10^7$ and with $B = 10^{12}$ G. Each line represents a different $\beta$ as marked on the panel.

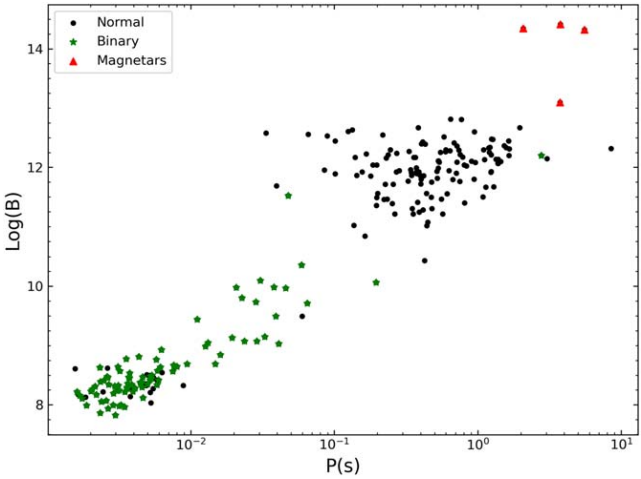

**Figure 8.** The magnetic field vs. the spin period for the observed pulsars shown in Figures 7.

When the neutrons in a superfluid vortex state make cyclotron motion, their rotational speed is $v_s(r) = \frac{n\hbar}{2m_n r}$. The energy is dissipated in the form of neutrinos and antineutrinos. According to Heisenberg's uncertainty principle, the core of the superfluid vortex should be normal neutron fluid that is rapidly rotating. The superfluid neutrons close to the vortex core gradually move outward. As a result, the normal neutrons located at the center of the superfluid vortex (with a higher density) also expand outward and will transform into superfluid neutrons during the motion of the superfluid vortex. The upper and lower bases of the neutron superfluid vortex filaments are pinned into the extremely neutron-rich atomic nuclei (with a lattice structure) in the neutron star crust. Once the normal neutron fluid in the core of the superfluid vortex filament spills out, the neutrons from the extremely neutron-rich nuclei in the shell flow continuously into the core of the superfluid vortex filament. It forms a kind of Ekman pump, in which the energy supply comes from the gravitational energy of the neutron star as it becomes denser.

For a budding neutron star, the initial superfluid quantum number ($n_0$) and the quantum state of the superfluid vortex filament are very high. The reason is detailed as follows. The matter in the collapsed core generated during a supernova explosion is in a highly turbulent state with many vortexes. The motion of the vortexes is very complex. The nascent neutron star has a radius of about 10 km and an angular velocity exceeding $1.0 \times 10^4\,\mathrm{s}^{-1}$, which corresponds to the minimum limiting spin period of 0.5 ms for a stable neutron star. Generally, during the gravitational collapse of the stellar core, (10–30)% of the angular momentum is transformed into vortexes of various sizes. Thus, the nascent neutron star will not break under the fast rotation. Soon after the birth, the nascent neutron star cools rapidly due to the intensive gravitational radiation and neutrino emission. A phase transition will occur and the neutron matter will become superfluid when the temperature drops to below the critical temperature of $T_c = \triangle/k_B \approx 2 \times 10^{10}$ K. Since there already exists highly chaotic vortex motion before the phase transition, the neutron superfluid should be in a superfluid vortex state. Note that the velocity circulation (vortex strength) is conserved for both the classical fluid vortex and the quantum superfluid vortex. Since the classical fluid vortex motion is highly turbulent, the subsequently produced superfluid vortex should also have a large quantum number $n_0$ (i.e., the quantum number of the superfluid vortex circulation). Nevertheless, the superfluid vortexes with a large quantum number are actually in an active state. They will slowly transit to states with smaller quantum numbers and release the angular momentum correspondingly. As a result, the rotation of the outer crust of the neutron star may be suddenly accelerated when the tension has accumulated to some extent. It may explain the glitches observed in young pulsars such as the Crab and the Vela pulsars (Peng et al. 2022). In addition, our model also predicts that the recoil velocity should be parallel to the neutron star's rotation axis. It is interesting to note that such a prediction is consistent with the observations of the Crab pulsar and the Vela pulsar (Lai et al. 2001).

We thank the anonymous referee for constructive suggestions that led to an overall improvement of this study. Z.L. is grateful to Nanjing University and Purple Mountain Observatory for valuable resources. This work was supported by National Natural Science Foundation of China (grant Nos. 11773062, 11873030, 12041306, and U1938201), the West Light Foundation of Chinese Academy of Sciences (2017-XBQNXZ-A-008), the National SKA Program of China No. 2020SKA0120300, the National Key R&D Program of China (2021YFA0718500), and by the science research grants from the China Manned Space Project with NO. CMS-CSST-2021-B11.

### ORCID iDs

Zheng Li  https://orcid.org/0000-0002-5755-7737
Xiang Liu  https://orcid.org/0000-0001-9815-2579
Yong-Feng Huang  https://orcid.org/0000-0001-7199-2906